# High-resolution and Fully-programmable Microwave-shaper

Jilong Li[1], Yitang Dai[1, *], Feifei Yin[1], Ming Li[2], and Kun Xu[1, 3]

1. State Key Laboratory of Information Photonics and Optical Communications, Beijing University of Posts and Telecommunications, Beijing, 100876, China

2. State Key Laboratory on Integrated Optoelectronics, Institute of Semiconductors, Chinese Academy of Sciences, Beijing 100083, China

3. School of Science, Beijing University of Posts and Telecommunications, Beijing, 100876, China

* Corresponding author: ytdai@bupt.edu.cn

**Abstract**

We here propose and demonstrate a point-by-point programmable broadband microwave spectrum processor with high-resolution up to tens of MHz. We achieve this by bandwidth-minified mapping a programmable optical spectrum processor, which has much larger bandwidth and lower frequency resolution, to a microwave one, and make sure the mapping is a similarity transformation. As a comparison, the traditional optical-to-microwave mapping based on super-heterodyne is bandwidth-preserved and identical. Here the optical spectrum processor is firstly sliced by a high-quality-factor optical resonator, which has periodic transmission peak with MHz-level bandwidth, and then is mapped to microwave domain by optical-frequency-comb-assisted multi-heterodyne. We demonstrate the high frequency resolution and full function reconfigurability experimentally. In a numerical example, we show that the group delay variation of optical spectrum processor could be greatly enlarged after mapping. The resolution improvement and group delay magnification distinguish our proposal significantly from previous spectrum mapping from optics to microwave.

**Introduction**

Reconfigurable radio frequency (RF) or microwave spectrum processing is fundamental in many fields such as wireless communication, radar, and electronic warfare [1-6]. In emerging information technologies, the instantaneous bandwidth to be processed could be as large as tens of GHz. Consequently, the flexibility of arbitrarily processing broadband signal becomes a must, adapting to future scenarios such as real-time channel equalization in wireless communications, matched filtering in frequency- or waveform-agile radar, simulating the target frequency response in deceptive jamming, and so on. For broadband spectrum-processing applications, microwave photonics (MWP) has received considerable attention due to the natural high carrier frequency of photonics [7-11]. Ultra-large bandwidth that is challenging to deal with in microwave domain are however relatively small when it is up-converted into optical domain, and then may be much easier to handle. By photonics, capacity of microwave filters, in terms of frequency tuning range and speed as well as function reconstruction, has been essentially improved [5, 12-16]. Currently, high processing resolution, fully-programmable, low processing latency, and possible integration for minimum size, weight, and power are highly desired in the field of microwave photonics filter (MPF).

There have been numerous demonstrations on MPFs, among which filters with finite impulse response (FIR) are most widely studied due to its superb function reconstruction [7-10]. In FIR scheme, values of to-be-processed signal from different time slots (which are commonly referred to as "taps") are weighted and summed together, and the target frequency response is proportional to the Fourier transform of weights. As a result, high-frequency-resolution reconfigurability covering large bandwidth requires both large number of programmable weights and long convolution time delay. The optically-incoherent FIR were firstly studied, and taps more than 100 has been supported by optical frequency comb (OFC) [5]. However, the incoherent weights are positive-only, and extra implementation, such as microwave photonic phase shift by stimulated Brillouin scattering (SBS) [17-19], nonuniformly spaced taps [20], and others [21, 22], are required to achieve the full reconfigurability. People have solved this problem by state-of-the-art optically-coherent MPF, seeking to build a programmable and integrated optical FIR filter and mapping it to RF domain by optical coherent detection [6, 23-30]. Limited by wafer size or propagation loss, integrated FIR itself shows difficulty in high frequency resolution (e.g. 100-MHz interference fringe requires 10-ns tap delay). Optical infinite impulse response (IIR) filters, which are typically optical resonators such as ring, Fabry–Pérot interferometer (FPI), phase-shifted Bragg grating, etc., are then introduced to increase the quality factor (Q-factor). Note the IIR filter has poor reconfigurability, and the capacity of fully-programmable frequency response by such combination has not been demonstrated both in theory and in experiment.

We notice that current coherent MPF employs *bandwidth-preserved mapping* from an optical spectrum processor to a microwave one. With a fixed frequency resolution, the Q factor of an optical filter should be four to five orders of magnitude higher than the target microwave. In order to release the implementation difficulty of such an optical processor, we here propose and demonstrate a *bandwidth-minified mapping* where a low-resolution optical spectrum processor is under similarity transformation while it is down-converted to microwave domain. In the novel MPF scheme, broadband RF signal is firstly up-converted and multicast by an OFC. All copies then pass through a high-Q-factor Vernier comb filter (VCF), of which the free spectrum range (FSR) is slightly different from the OFC. The spectrum of RF signal is then sliced, and each part is separated far from its neighbors due to the large FSR of VCF, which can then be easily and independently spectrum-processed by a low-resolution programmable optical filter. By multi-heterodyne, all slices are down-converted to RF domain to assemble the target output microwave spectrum. The above MPF scheme is demonstrated by experiment and simulation in this paper.

**MPF design and theory**

In FIR-based MPF, each tap is individually designed in time domain to assemble the impulse response which is inverse Fourier transform of any target frequency response, $H_{\mathrm{MPF}}(\Omega)$. Here on the contrary, $H_{\mathrm{MPF}}(\Omega)$ is assembled in frequency domain from a series of high-Q-factor optical resonators, $H_{\text{OR}}(\Omega - \Omega_k)$, of which the center frequency, $\Omega_k$, ranges the target bandwidth, and the loss and phase shift, $|\alpha_k|e^{i\arg(\alpha_k)}$, are individually designed before assembly. Mathematically, we approach $H_{\text{MPF}}(\Omega)$ by $\sum_k \alpha_k H_{\text{OR}}(\Omega - \Omega_k)$. Note MPF by stitching a few independent optical resonators has been previously demonstrated in [31]. However, since all high-Q-factor responses should be aligned precisely together with small frequency tolerance (which may be around tens of MHz or even less) as well as phase tolerance (otherwise the assembly is unstable or complicated feedback control of each resonator is required), such direct implementation is difficult and limits the numbers of resonators to be assembled. Here we obtain precisely stitching through an optical frequency comb and a comb filter where the both frequency spacings are strictly equal respectively. The proposed MPF scheme is shown in Fig. 1.

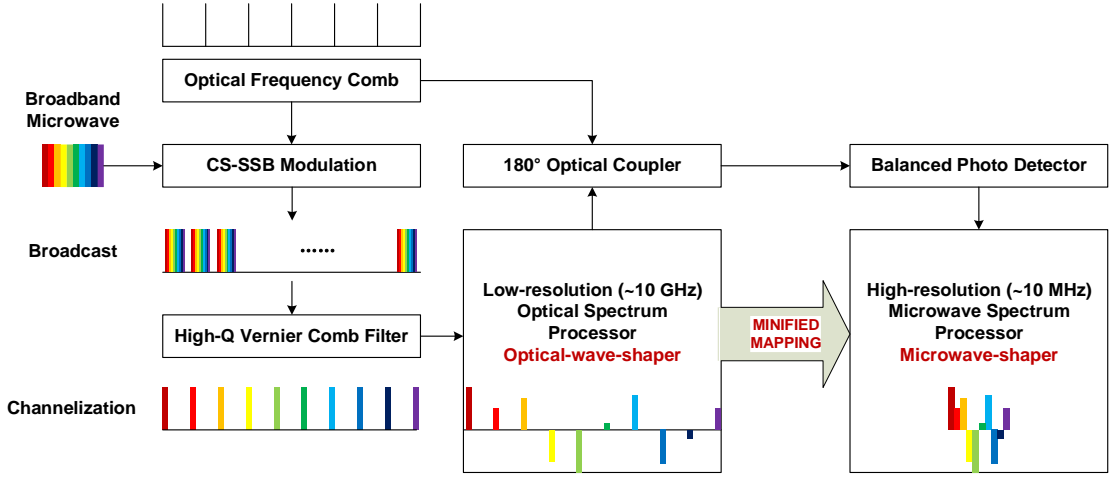

Figure 1. The proposed microwave-shaper by bandwidth- minified-mapping an optical wave-shaper

The input broadband microwave signal is firstly broadcast by OFC where $\text{FSR}_{\text{OFC}}$ is larger than twice of the highest frequency within signal bandwidth. Carrier-suppression single-sideband (CS-SSB) modulation is employed. The broadcast spectrum is sliced by a VCF where its $\text{FSR}_{\text{VCF}}$ is slightly different from $\text{FSR}_{\text{OFC}}$. The VCF could be a high-Q-factor optical resonator, and one can find the same ultra-narrow bandpass filtering, $H_{\text{OR}}[\omega - (\omega_{\text{VCF}}^0 + 2k\pi\text{FSR}_{\text{VCF}})]$, in each FSR. Here $\omega_{\text{VCF}}^0$ is the center frequency of its 0th channel. We assume frequency of the 0th line of OFC, $\omega_{\text{OFC}}^0$, is less than $\omega_{\text{VCF}}^0$, $\text{FSR}_{\text{OFC}} < \text{FSR}_{\text{VCF}}$, and the upper-sideband is selected after CS-SSB modulation. Different conditions will result in similar conclusion. Within the $k$th channel which contains $k$th line of OFC and $k$th transmission peak of VCF, spectrum slice of microwave signal around $\Omega_k = (\omega_{\text{VCF}}^0 - \omega_{\text{OFC}}^0) + 2k\pi(\text{FSR}_{\text{VCF}} - \text{FSR}_{\text{OFC}})$ is extracted. Then the slice is weighted with complex $\alpha_k$ by the following optical spectrum processor, and is down-converted to microwave domain by the same $k$th comb line. Within the $k$th channel, the physics is the same as traditional optically-coherent MPF, which results in microwave frequency response of $\alpha_k H_{\text{OR}}(\Omega - \Omega_k)$. In Fig. 1 the multi-heterodyne is used instead of traditional super-heterodyne, and all slices are down-converted simultaneously so that frequency responses of all channels are stitched together and one can finally get

$$H_{\text{MPF}}(\Omega) \propto \sum_k \alpha_k H_{\text{OR}}\left(\Omega - \left(\omega_{\text{VCF}}^0 - \omega_{\text{OFC}}^0\right) - 2k\pi(\text{FSR}_{\text{VCF}} - \text{FSR}_{\text{OFC}})\right) \qquad (1)$$

Two issues should be considered for a correct Eq. (1). Firstly, in the multi-heterodyne, any down-conversion product between sliced spectrum and OFC line from different channels should be out of the bandwidth of microwave signal. A sufficient condition is that $\text{FSR}_{\text{OFC}}$ is larger than twice of the highest frequency of input signal. This condition is unnecessary, however. For example, when input microwave signal is limited within $(\text{FSR}_{\text{OFC}}/2, \text{FSR}_{\text{OFC}})$, Eq. (1) is still correct. Secondly, besides the optical spectrum processor, $\alpha_k$ may be impacted by other factors. Usually, $H_{\text{OR}}$ is identical for every channel, but OFC has significant non-uniformity both in power and in phase. The additional unflattened amplitude response among each channel could be equivalently accounted for by the optical spectrum processor; in other words, the amplitude non-uniformity induced by OFC could be corrected by the programmable optical processor. The phase difference among OFC lines is automatically eliminated since up- and down-conversion in each channel employs the same optical local oscillation (LO). Accordingly, the phases of OFC lines have no contribution to $\alpha_k$. Besides, all LO lines pass through a common optical path, so do the spectrum slices. As a result, the possible phase fluctuations due to propagation are the same for all channels, which may introduce an uncertain but uniform phase shift to all $\alpha_k$. We can conclude that assembly of all high-Q-factor resonate responses by our proposal is stable in phase, and complex weight, $\alpha_k$, in each channel can be fully controlled by the optical spectrum processor.

The proposed point-by-point programmable microwave spectrum stitching has obvious advantages over previous reports. Firstly, all slices are precisely aligned in frequency, which is spaced by $\text{FSR}_{\text{VCF}} - \text{FSR}_{\text{OFC}}$ and is ensured by strictly-equally-spaced OFC and VCF. Frequency tuning could be obtained by changing $\omega_{\text{VCF}}^0 - \omega_{\text{OFC}}^0$. Secondly, all resonator responses ($H_{\text{OR}}$) are from a single device, so that they could be highly uniform with good performance. Meanwhile, the number of responses to be assembled could be quite large since the current OFC and optical resonator are easily broadband. Thirdly, the complex weight, $\alpha_k$, is set by programmable optical processor which is current commercial available such as Waveshaper from Finisar Corporation [32]. The integrated version has also been demonstrated by [33, 34]. Note that the frequency resolution of optical processor, which is $\text{FSR}_{\text{VCF}}$ since $\alpha_k$ is set every $\text{FSR}_{\text{VCF}}$, is much lower than $\text{FSR}_{\text{VCF}} - \text{FSR}_{\text{OFC}}$, while the latter is the resolution of proposed MPF since the frequency response changes every $\text{FSR}_{\text{VCF}} - \text{FSR}_{\text{OFC}}$. The resolution is greatly improved by a factor of

$$M = \frac{\text{FSR}_{\text{VCF}}}{\text{FSR}_{\text{VCF}} - \text{FSR}_{\text{OFC}}} \qquad (2)$$

after optics-to-microwave mapping. In physics, after the OFC-based broadcast and VCF-based slicing, the bandwidth of input microwave signal is equivalently magnified by factor of $M$, so that the target high-resolution spectrum processing could be achieved by a low-resolution one. As can be seen, Fig. 1 shows a bandwidth-minified mapping from an "optical-wave shaper" to a "microwave shaper" rather than the traditional bandwidth-preserved one, and the bandwidth compression ratio is also $M$. The mapping can be expressed approximately by

$$H_{\text{MPF}}\left(\left(\omega_{\text{VCF}}^0 - \omega_{\text{OFC}}^0\right) + \Omega\right) \propto \alpha\left(\omega_{\text{VCF}}^0 + M\Omega\right) \qquad (3)$$

where $\alpha(\omega)$ is the frequency response of optical spectrum processor. Bandwidth of state-of-the-art optical spectrum processor (such as Waveshaper from Finisar Corporation) could be as broad as several THz, which supports large bandwidth compression as well as resolution improvement.

**A proof-of-concept experiment**

Our experiment setup follows Fig. 1. The OFC is generated by phase modulating a continuous-wave (CW) light with $\text{FSR}_{\text{OFC}} \approx 10$ GHz sinusoidal wave, as shown in Fig. 2 (a). The CW light is a fiber laser (from AOI) at 193.43 THz, with linewidth about 30 kHz. As the number of comb lines is determined by the strength of phase modulation (PM), the driving power is set as large as possible (the total modulation depth is about 13). Two phase shifters before PMs are used for synchronization. The comb is shaped and flattened by a Mach-Zehnder modulator (MZM), which is driven by around 5 GHz sinusoidal wave (which is synchronized with the above 10 GHz wave) and biased at maximum transmission point. The generated OFC ranges from 193.31 THz to 193.57 THz as shown in Fig. 2 (b). 26 lines are obtained within the 5-dB bandwidth.

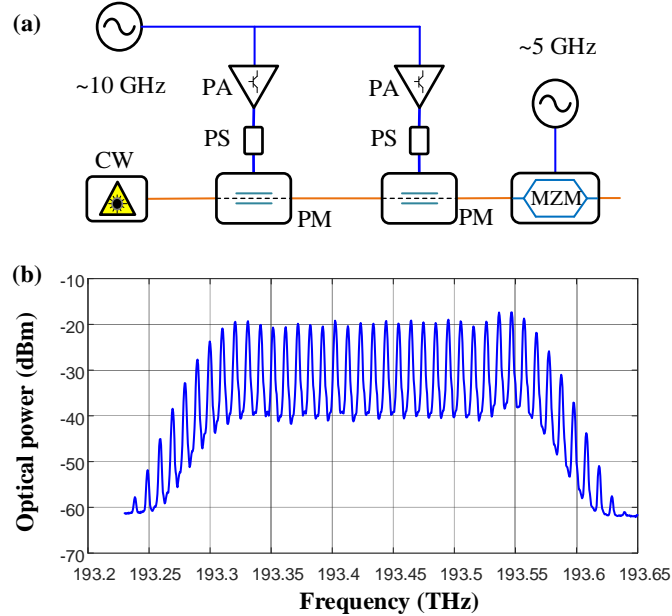

Figure 2. (a) OFC generator. PA: power amplifier. PS: phase shifter. (b) Measured OFC spectrum.

The VCF is a fiber FPI (from Micro-Optics), its $\text{FSR}_{\text{VCF}}$ is about 10.2 GHz, and the 3 dB bandwidth of each transmission peak ($H_{\text{OR}}$) is about 50 MHz. We control the frequency of CW laser to tune $\omega_{\text{VCF}}^0 - \omega_{\text{OFC}}^0$. A waveshaper (from Finisar) is used as the programmable optical spectrum processor, and the resolution is 10 GHz. A vector network analyzer (VNA, from Keysight Technologies, Inc) is used to measure the S21 parameter of the MPF.

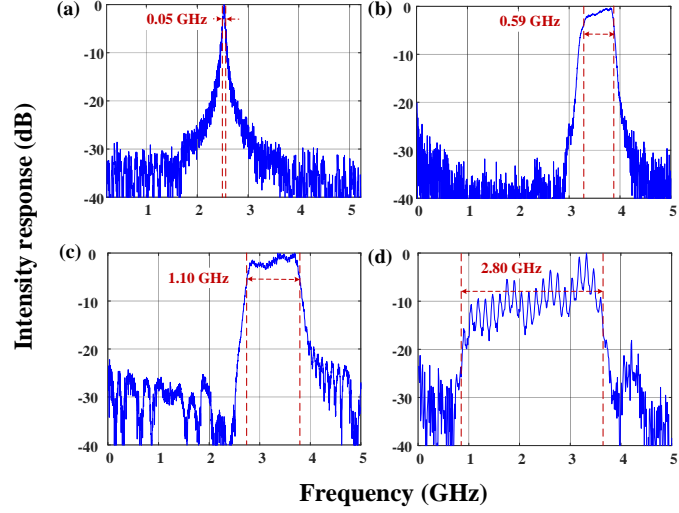

Figure 3. Intensity response of the proposed MPF when $\alpha_k \equiv 1$ and $\text{FSR}_{\text{VCF}} - \text{FSR}_{\text{OFC}}$ are (a) 0 MHz, (b) 22 MHz, (c) 42 MHz, and (d) 142 MHz, respectively. In (a) to (c) the OFC is shown in Fig. 2(b) and the number of lines is around 26; in (d) the number of lines is around 20, otherwise the bandwidth of MPF is larger than $\text{FSR}_{\text{OFC}}/2$.

Firstly, the superposition of all slices during multi-heterodyne is observed by tuning $\text{FSR}_{\text{VCF}} - \text{FSR}_{\text{OFC}}$ while $\alpha_k \equiv 1$. Figure 3(a) to 3(d) show intensity responses, $|H_{\text{MPF}}|^2$, when FSR differences are around 0, 22, 42, and 142 MHz, respectively. If two FSRs are the same, all slices correspond to the same spectrum position of microwave signal, and $H_{\text{MPF}} \propto H_{\text{OR}}$ according to Eq. (1). The Lorentzian peak with 3 dB bandwidth of 50 MHz, shown in Fig. 3(a), is consistent with theory. When FSR difference is much larger than bandwidth of $H_{\text{OR}}$, one can see separated transmission peak within $H_{\text{MPF}}$, as shown in Fig. 3(d), and every peak is proportional to $H_{\text{OR}}$. Continuous intensity response can be found when $\text{FSR}_{\text{VCF}} - \text{FSR}_{\text{OFC}}$ is approximately the same as or less than the bandwidth of $H_{\text{OR}}$, as shown in Figs. 3(b) and 3(c). Under limited number of OFC lines ($N_{\text{OFC lines}}$), the bandwidth of $H_{\text{MPF}}$ is obviously $N_{\text{OFC lines}} \cdot (\text{FSR}_{\text{VCF}} - \text{FSR}_{\text{OFC}})$. The four bandwidths in Fig. 3 agree well with theory, which provides a simple way for bandwidth configuration. Note the continuous intensity responses in Figs. 3(b) and 3(c) also confirm the stable in-phase superposition during multi-heterodyne.

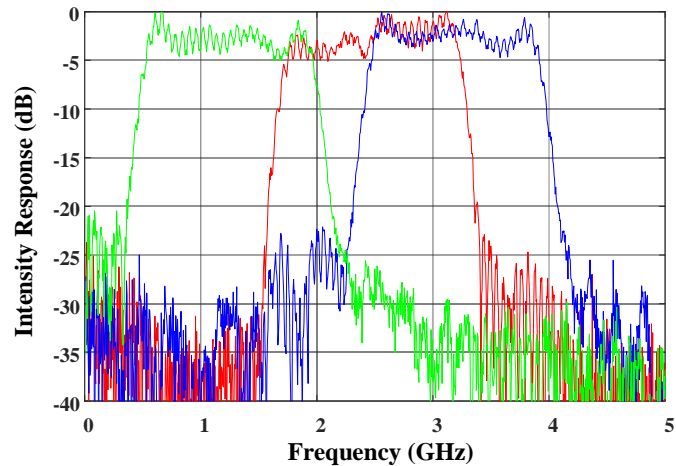

Figure 4. Frequency tuning of the proposed MPF by changing laser frequency.

Secondly, the frequency tuning of the proposed MPF is demonstrated by changing laser frequency, i.e. $\omega_{VCF}^0 - \omega_{OFC}^0$. In Fig. 4, three bandpass filters range from 0.5 GHz to 2 GHz, 1.75 GHz to 3.20 GHz and 2.50 GHz to 3.92 GHz, respectively. $FSR_{VCF} - FSR_{OFC} = 60$ MHz so that 3 dB bandwidths are all 1.5 GHz. The bandwidth compression ratio is 176. The fluctuation in the passband is below 5 dB, which is mainly induced by the flatness of the OFC. The sideband suppression ratio is about 20 dB, which we believe suffers from the limited extinction ratio of the fiber FPI.

Thirdly, the function reconfiguration is demonstrated in Figs. 5(a) and 5(b), where notch filter and bandpass filter with a slope intensity response are shown, respectively. $FSR_{VCF} - FSR_{OFC} = 60$ MHz, which is the same as bandpass filter in Fig. 4. As shown in Eq. (3), such functions can be easily obtained by setting the same but bandwidth-enlarged and low-resolution functions on optical waveshaper. In Fig. 5(a), notches are located in 2.3 GHz, 2.6 GHz, and 2.83 GHz, respectively, which all shows sharp feature around tens of MHz. The narrowest passband shown in Fig. 3(a) as well as the notches in Fig. 5(a) shows the resolution of our MPF, which is determined by the VCF (i.e. $H_{OR}$). The capacity of high-resolution, point-by-point response definition is then feasible by our proposal.

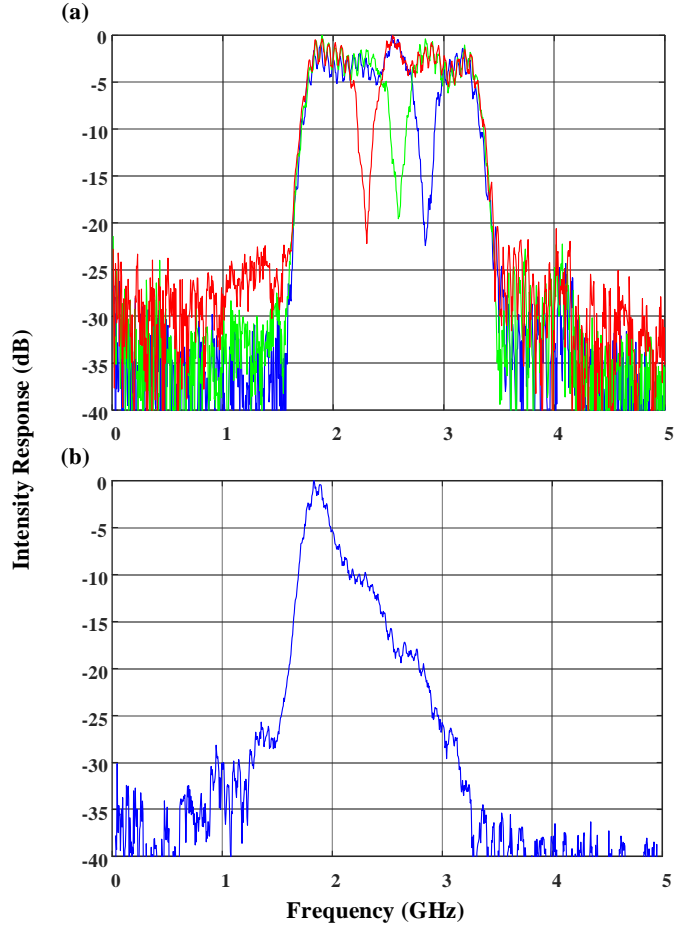

Figure 5. (a) Notch filters at different frequencies. (b) Bandpass filter with a slope intensity response.

# Enlarged time delay response – a numerical example

The phase response property of the proposed MPF is demonstrated by simulation. According to Eqs. (1) and (3), though the response bandwidth of spectrum processor is compressed by a factor of $M$ during mapping from optics to microwave, the weight, $\alpha_k$, at each point is preserved, so as to its phase. Note it takes frequency interval of $\text{FSR}_\text{VCF}$ if phase response changes from $\alpha_k$ to $\alpha_{k+1}$. After mapping the required frequency interval is compressed to $\text{FSR}_\text{VCF} - \text{FSR}_\text{OFC}$. Since the group delay is proportional to resulted phase variation divided by frequency interval, the group delay spectrum, $\tau$, after the proposed bandwidth-minified mapping will be enlarged by the same factor of $M$ in magnitude,

$$\tau_\text{MPF}\left(\left(\omega_\text{VCF}^0 - \omega_\text{OFC}^0\right) + \Omega\right) \approx M \cdot \tau_\alpha(\omega_\text{VCF}^0 + M\Omega) \qquad (4)$$

In the following numerical example, $\text{FSR}_\text{OFC} = 20$ GHz and $N_\text{OFC lines} = 51$. All comb lines have uniform intensity. $\text{FSR}_\text{VCF} = 20.05$ GHz, and the VCF is an ideal FPI where bandwidth of each resonator peak is 50 MHz. The frequency response of FPI is $S_\text{FPI}^{21} = (1-r^2)/\left(e^{-i\theta/2} - r^2 e^{i\theta/2}\right)$ where $r$ is the reflectivity of each end, $\theta = \omega/\text{FSR}_\text{VCF}$, and $\omega$ is the angular frequency of input lightwave. Accordingly, the programmable bandwidth of MPF is about 2.5 GHz, which is centered at 5 GHz by setting proper $\omega_\text{VCF}^0 - \omega_\text{OFC}^0$. The bandwidth compression ratio is $M = 401$ according to Eq. (2). The optical spectrum processor has uniform intensity response and constant time delay, $\tau_\alpha$, over the bandwidth, that is, $\alpha(\omega_\text{VCF}^0 + \omega) = e^{i\omega\tau_\alpha}$. The calculated intensity response under $\tau_\alpha = 0$ and phase responses under different $\tau_\alpha$ are shown in Figs. 6(a) and 6(b), respectively.

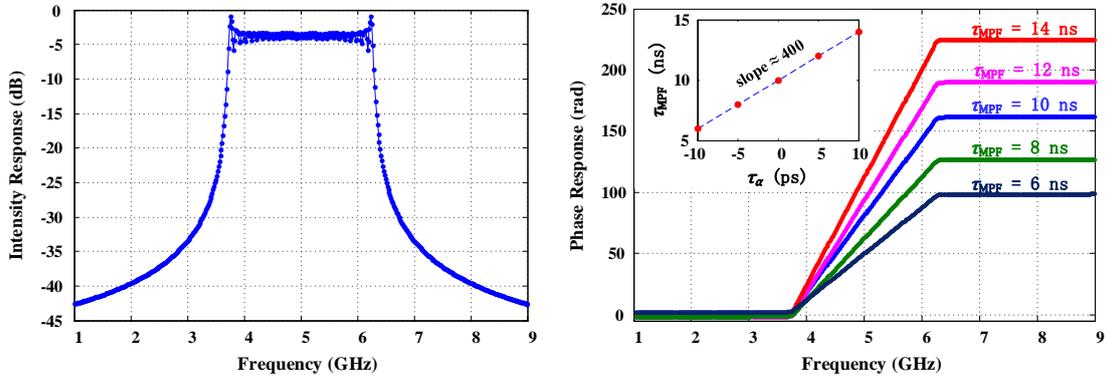

Figure 6. (a) Simulated intensity response when $\tau_\alpha = 0$. (b) Simulated phase response when group delay of optical processor is -10, -5, 0, 5, and 10 ps, respectively. Inset: group delay variation of MPF is enlarged greatly compared with optical processor.

In Fig. 6(a) one can observe sharp edges of the bandpass filter, similar with experiment result in Fig. 4. Within its 2.5 GHz passband, we can see linear phase variation along with frequency, which shows group delay of $\tau_\text{MPF} = 10$ ns when $\tau_\alpha = 0$. Simulation shows that tiny variation of $\tau_\alpha$ results in large $\tau_\text{MPF}$ variation. When $\tau_\alpha$ changes from -10 to 10 ps, the group delay of MPF changes from 6 ns to 14 ns, that is, the group delay variation is enlarged by 400. The number is consistent with Eq. (4). Such group delay magnification will be very useful for key applications, including phase array antenna, compressive receiver, etc., where large time-bandwidth product or fast tuning is desired.

**Conclusion**

In this paper we experimentally demonstrated a high-frequency-resolution microwave photonics filter with full function reconfigurability. With the same setup, MPF frequency responses of rectangle-like with different bandwidth, triangle-like, notch-type with different location, as well as ultra-narrow bandpass shapes were realized. Frequency tunability was also reported. The obtained frequency resolution was around 60 MHz. The proposed MPF was achieved by bandwidth-minified mapping a programmable optical spectrum processor to a microwave one. Different from traditional mapping which is based on single-frequency lightwave and super-heterodyne, here we realized the novel mapping by optical frequency comb, a Vernier comb filter, as well as multi-heterodyne. The bandwidth compression during mapping results in greatly increased frequency resolution as well as group delay magnification. We would like to note that our proposal is totally different from programmable FIR scheme where array of true delay lines is required, and is compact and ready for integration.